\documentclass[lettersize,journal]{IEEEtran}

\usepackage{amsmath,amsfonts}

\usepackage{array}
\usepackage{makecell}
\usepackage{alphabeta}
\usepackage{multirow}
\usepackage{mdwmath}
\usepackage{mdwtab}
\usepackage{eqparbox}
\usepackage[caption=false,font=normalsize,labelfont=sf,textfont=sf]{subfig}
\def\BibTeX{{\rm B\kern-.05em{\sc i\kern-.025em b}\kern-.08em
    T\kern-.1667em\lower.7ex\hbox{E}\kern-.125emX}}
\usepackage{textcomp}
\usepackage{stfloats}
\usepackage{url}
\usepackage{verbatim}
\usepackage{graphicx}
\usepackage{cite}
\usepackage{upgreek}
\usepackage{algorithm}
\usepackage{algorithmic}
 
\usepackage{bm}
\usepackage{balance}
\usepackage{makecell}

\begin{document}

\title{Topology Design of Reconfigurable Intelligent Surfaces Based on Current Distribution and Otsu Image Segmentation}

\author{Zhen Zhang, \IEEEmembership{Member, IEEE}, Jun Wei Zhang, Hui Dong Li, Junhui Qiu, Lijie Wu, Wan Wan Cao, Ren Wang, Jia Nan Zhang, \IEEEmembership{Member, IEEE}, and Qiang Cheng, \IEEEmembership{Senior Member, IEEE}     % <-this % stops a space
\thanks{Manuscript received **; revised **, 2025. This work is supported by the National Key Research and Development Program of China (2017YFA0700201, 2017YFA0700202, 2017YFA0700203, 2021YFA1401002, 2023YFB3811503), the National Natural Science Foundation of China (62171124, 61631007, 61571117, 61138001, 61371035, 61722106, 61731010, 11227904, 62071211), the 111 Project (111-2-05), Natural Science Foundation of Jiangsu Province (BK20212002), Zhejiang Provincial Natural Science Foundation of China(LQ23F040009), and the State Key Laboratory of Millimeter Waves, Southeast University (K202403). (Corresponding author: Qiang Cheng ). % <-this % stops a space

Zhen Zhang is with School of Electronic and Communication Engineering, Guangzhou University, China, and also with the State Key Laboratory of Millimeter Waves, Southeast University, Nanjing 210096, China (zhangzhen@gzhu.edu.cn) . 

Junhui Qiu with School of Electronic and Communication Engineering, Guangzhou University, China(e-mail: qiujunhui@e.gzhu.edu.cn). 

Ren Wang is with the Institute of Applied Physics, University of Electronic Science and Technology of China, Chengdu 611731, China (e-mail: rwang@uestc.edu.cn).

Jun Wei Zhang, Lijie Wu, Hui Dong Li, Jia Nan Zhang and Qiang Cheng are with the State Key Laboratory of Millimeter Wave, Southeast University, Nanjing 210096, China (email: jiananzhang@seu.edu.cn, qiangcheng@seu.edu.cn).}}


\maketitle

\begin{abstract}

Miniaturization of reconfigurable intelligent surface (RIS) elements is a crucial trend in the development of RISs. It not only facilitates the attainment of multifunctional integration but also promotes seamless amalgamation with other elements. The current on the RIS element plays a crucial role in determining the characteristics of the induced electromagnetic field components. Segments with high current intensity determine the performance of RIS elements. Carving the parts with strong current distribution density into the metal patch of RIS element structure can achieve miniaturization. Based on this insight, this work proposes a topology design method that leverages current distribution and image processing techniques to achieve efficient miniaturization of the RIS elements. In this proposed method, we first obtain the current distribution across different operational states and the period of the working frequency. Next, we employ the Otsu image segmentation method to extract relevant image information from the current distribution images of the RIS elements. Subsequently, we utilize linear mapping techniques to convert this image information into the structure of RIS elements. Then, based on the structure of the RIS elements, the Quasi-Newton optimization algorithm is utilized to obtain the parameters of the tunable device that correspond to various operational states. As a result, we successfully construct the structural topology of the RIS elements based on their current distribution, designing areas with strong current distribution as metal patches. To validate the performance of the proposed method, a $16\times 16$ 3-bit RIS was developed, fabricated and measured. Compared with existing RIS designs, the proportion of the top-layer metal patches is smaller, which provides the possibility for integrating other functions and devices.  

\end{abstract}


\section{Introduction}
\IEEEPARstart
{R} {econfigurable}  Intelligent Surface (RIS), as a programmable electromagnetic surface, is capable of dynamically manipulating the wireless signal propagation environment through the integration of numerous controllable RIS elements \cite{r1,r2,rrr3}.
Specifically, RIS enables some attractive functionalities such as beamforming, directional control, focusing, and scattering adjustment, which can effectively optimize wireless communication channels and significantly improve wireless communication quality\cite{rr1,rr4,rrr3}. Currently, RIS has been extensively applied in various fields. In terms of signal coverage, it has successfully eliminated coverage dead zones, significantly expanding the reach of cellular signals\cite{rrr4,rrr5}. In the realm of secure communication, RIS intelligently adjusts the phase of reflected signals to cancel out direct signals, thus markedly reducing the risk of information leakage\cite{r8}. Consequently, RIS is gradually emerging as a hot topic in the microwave research community.

% =======
% FIG. 1
% =======
\begin{figure}
\centering
\subfloat[]{\includegraphics[width=3.2in]{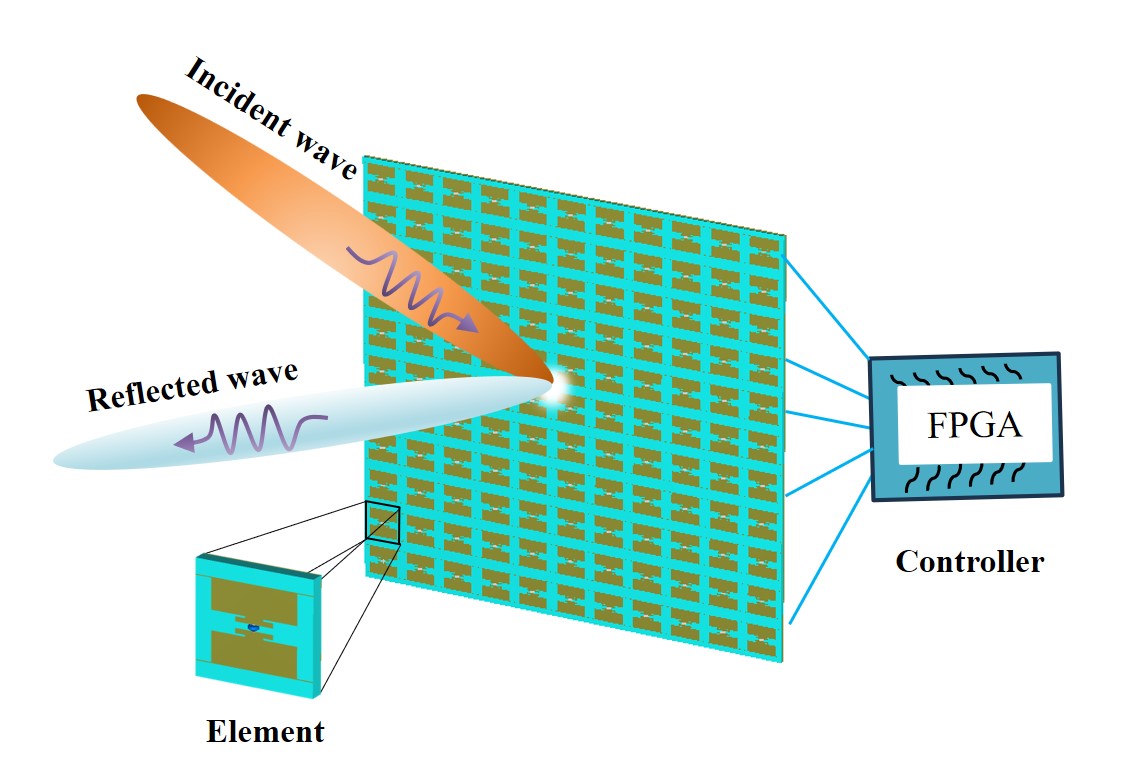}}

\subfloat[]{\includegraphics[width=3in]{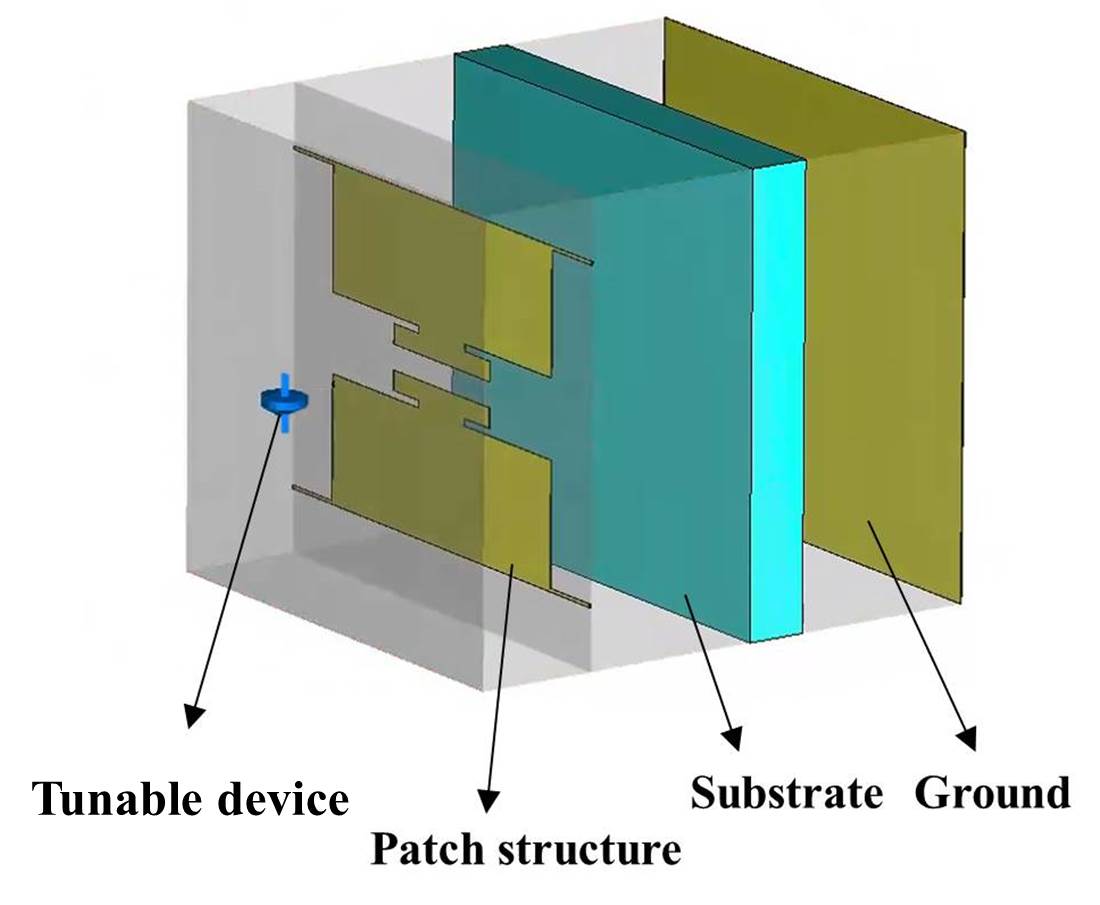}}
\caption{llustrative diagram of the RISs, (a) array, and (b) element. }
\end{figure}

The design of the RIS system is crucial for implementing several expected and diverse functions \cite{rr25,rr26}. The RIS system comprises two parts: the RIS element and the array control module, as depicted in Fig. 1(a). Each element is embedded with some tunable devices (such as p-i-n (PIN) diodes and varactor diodes), as shown in Fig. 1(b). These devices alter the response of the RIS element by toggling the operating state of the tunable devices. The array control module is constituted by digital signal processors such as microcontrollers (MCU) and field-programmable gate arrays (FPGA). These processors are capable of switching the operating state of the tunable devices in real time in a programmable way, thereby regulating the intelligent surface's response to wireless signals. The design approach for RIS is of great significance for its real-world applications  \cite{r8,r9,rr8,rr9}.

Recently, significant progress has been made in the design of RIS elements, which mainly diverges into two main areas: parameter optimization \cite{rr10,r28,r18} and topology optimization\cite{rr11, rr12, r13}. Parameter optimization involves adjusting the parameters of an existing structure, such as length and width. In contrast, topology optimization aims to generate a new topological structure, which offers a higher degree of design freedom.  This work focuses on the topology optimization of RISs. Currently, two main approaches dominate the topology optimization algorithms for RISs: one is based on machine learning \cite{rr11, rr12, r13, rrr12}, and the other is based on a multi-port network model \cite{r34,rr9}.  The machine learning-based topology optimization utilizes datasets from electromagnetic (EM) simulations to train the machine learning model\cite{r11}. Subsequently, this model is applied to obtain the topological design of RIS elements that meet specific design criteria. The main advantage of this method lies in its strong data processing capability, enabling the rapid identification of complex relationships between topological designs and EM responses. However, it still relies on EM simulation samples as these samples are essential for the modeling process.
Another approach involves a multi-port network-based design for RIS \cite{rrr11}. In this method, the RIS element is divided into several rectangular metallic patches connected by internal ports. By analyzing the multi-port parameters, a connection is established between the loads at the internal ports and the reflection coefficients of the RIS elements. Adjusting these loads can result in changes in the topological current of the RIS. This optimization technique offers advantages such as decreased reliance on EM simulations, shorter design time, and the ability to alter the current topology. Nevertheless, the RIS structures generated by this method often resemble the original design, making it challenging to achieve miniaturization. 

The miniaturization of RIS elements represents a significant and essential trend within the progression of RISs. Not only the design facilitates the attainment of multifunctional integration, it also promotes seamless amalgamation with other elements.
To achieve the miniaturization of the metasurface antenna, a symmetrical quasi-fractal slot is loaded onto each square metal patch in \cite{r37}. In \cite{r38}, miniaturized wideband metasurface antennas are realized by cross-layer capacitive loading. Furthermore, miniaturization of metasurface antennas is accomplished using characteristic mode analysis as described in \cite{r39}.  For RISs, there are relatively few designs for the miniaturization of structures.

The distribution of electrical current is a crucial factor in the design of RIS elements, as it has a significant influence on their operational efficiency and overall performance. Therefore, carving parts with strong current distribution density as the top layer metal patch structure of RIS elements can achieve a natural optimal topology and realize the miniaturization of the structure.  
Based on the above analysis, in this work, we propose integrating current distribution considerations into the topology optimization of RIS elements with the intention of enabling rapid and precise topological design. Firstly, we extract the current distribution maps of RIS elements under different working states and at various moments. Secondly, image processing is carried out on the current distribution maps to extract the information with stronger current distribution density. Finally, the parts with stronger current distribution density are mapped into the top-layer metal patch structure of RIS elements for achieving miniaturization of the structure. The specific contributions of this work are outlined as follows:

(1) A new method for designing topologies of RIS based on current distribution is introduced. This method incorporates image processing technology into RIS design; in other words, it uses current distribution as a link to enhance the topological structure. To our knowledge, this is the first approach that exploits current distribution graphs for the automatic design of RIS topology.

(2) A precise mapping technique from 2D images to RIS structures has been developed. This technique ensures an accurate relationship between images and structures while considering current distribution characteristics and physical structural limitations, thus ensuring the rationality of RIS topology design.

(3) By leveraging the aforementioned theories and methodologies, this work successfully design and fabricates a 3-bit phase-modulated RIS. The miniaturization ratio of the RIS is 0.86, which is far lower than that of the existing designs, indicating that this design is highly compact. 

The remainder of this work is structured as follows: Section II provides the details of the proposed topology design method. Section III presents the application and results of the proposed method. Finally, Section IV concludes the work.

% =======
% FIG. 2
% =======
\begin{figure*}[ht]
\centering
\includegraphics[width=7in]{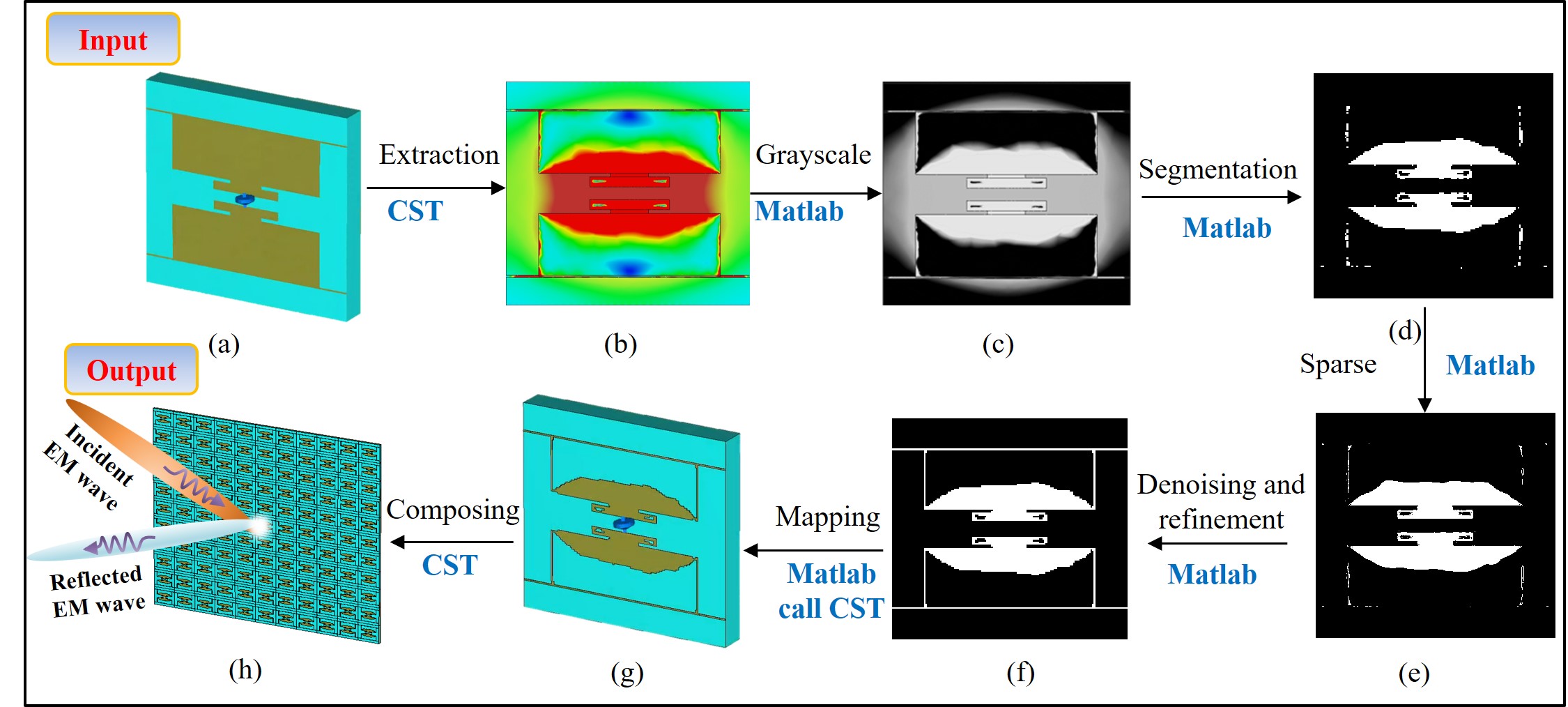}
\caption{Flowchart of the topology design based on the current distribution of RISs. (a) initial structure, (b) current distribution, (c) grayscale image, (d) segmented image, (e) sparse image, (f) image after denoising and refinement of the feeding structure, (g) RIS element, and (h) RIS system. }
\end{figure*}

\section{The Proposed Topology Design Method }

In the design of RISs, performance optimization and miniaturization are two crucial indicators. Performance optimization aims to enhance the EM wave manipulation capabilities of RISs. Miniaturization, on the other hand, is to meet the requirements of modern wireless communication systems in terms of device size and weight. Based on this, we formulate the design problem of RISs as a constrained optimization problem in this study, as shown below: 

\begin{equation}
\left\{\begin{array}{cc}\min & f(\boldsymbol{x}) \\\text { s.t. } & g(\boldsymbol{x}) \leq 0  \\\end{array}\right.
\end{equation}
where $\boldsymbol{x}$ is the design variables, $f(\boldsymbol{x})$ is the objective function, and it is usually composed of the phase and amplitude of the reflection characteristic for RIS elements. $ g(\boldsymbol{x})$ is the nonlinear constraint functions related to the miniaturization of RIS elements. 

For RIS design, the current distribution plays a crucial role in determining the characteristics of the induced field components. Hence, areas with higher current density have a more significant impact on the performance of the RIS element. It can be concluded that minimizing or eliminating patches with lower currents has little effect on the overall performance. Therefore, this approach brings considerable benefits to the miniaturization of RISs.

Typically, the structure of an RIS element consists of a top metal patch, tunable devices, a substrate, and a ground, as illustrated in Fig. 1(a). In this work, we propose a new method for designing the topology of the top metal patch on RIS elements. The topology design process is illustrated in Fig. 2. Fig. 2 outlines the transition from a colored 2D current distribution map to a grayscale image, followed by image segmentation, sparsification, and the mapping of image matrix to the topology structure, which leads to the performance optimization of the elements and the array composition. 
The grayscale representation of the current distribution of the RIS element is discussed in Section II-A. Next, we apply image processing techniques to extract the relevant information from the current distributions, as explained in Section III-B. Then, we use the extracted image data to create the topological configuration of the RIS elements, as detailed in Section III-C. Finally, we utilize the optimization algorithm to obtain RIS elements in different working states, as outlined in Section III-D.

\subsection {Grayscale approach for current distribution images}

Consider a RlS element with an initial EM structure depicted in Fig.2(a), based on this existing structure, we employ EM simulation to acquire the current distribution image, as shown in Fig. 2(b). Notably, these current distribution images are color pictures possessing three channels. Specifically, the color current distribution image $\boldsymbol{D}$ is constituted by three component matrices: $\boldsymbol{R}$, $\boldsymbol{G}$, and $\boldsymbol{B}$, with the pixel values of each channel matrix ranging from 0 to 255. Image grayscale processing refers to the process of converting the pixel information of a color image into a grayscale image containing only luminance information through specific algorithms \cite{r40}. The data volume of a grayscale image is significantly reduced compared to that of a color image. 
In this work, the grayscale method is employed to extract the image information. Specifically, during the grayscale process, we have prior knowledge that the red component signifies the intensity of the current for the RIS elements. We consequently select the information in the red channel as the grayscale processed image as follows:

\begin{equation}
\boldsymbol{G}_{r} = \boldsymbol{R}
\end{equation}
where $\bm {G}_{r}$ is the grayscale image matrix.

The current distribution of the RIS element changes with the period of the working frequency. Moreover, the RIS current distributions corresponding to different operating states of tunable devices are also different. The relationship between current distribution and moment and working state is defined as $\boldsymbol{G}_{r}$($t$, $s$), where $t$ represents the moment and $s$ denotes a specific working state, as shown in Fig. 3. When designing RIS, we need to comprehensively consider and optimize its performance across various states and moments. So, we obtain the maximum of current distributions across possible states and moments of each pixel of the current distribution matrix, which is defined by the following: 

\begin{equation}
{D}({i,j} )= {\textstyle Max_{t=1}^{T}Max_{s=1}^{S}{G}_{r}^{i,j}({t,s})}
\end{equation}
where ${G}_{r}^{i,j}({t,s})$ is the pixel of matrix $\boldsymbol{G}_{r}({t,s})$ in the $i$-th row and $j$-th column.  $i$ ranges from 1 to $I$ and $j$ ranges from 1 to $J$. Here, $I$ and $J$ denote the number of rows and columns of the current distribution matrix, respectively. $T$ is the number of considered moments, and $S$ is the number of considered working states.

 By traversing all pixels of the current distribution matrix $\boldsymbol{G}_{r}$ using Eq. (3), we can obtain the maximum current distribution matrix $\boldsymbol{D}$ for all moments and working states of the RIS elements.

% ==== FIG3
\begin{figure}%[!htbp]
  \centering
  \includegraphics[scale=0.27]{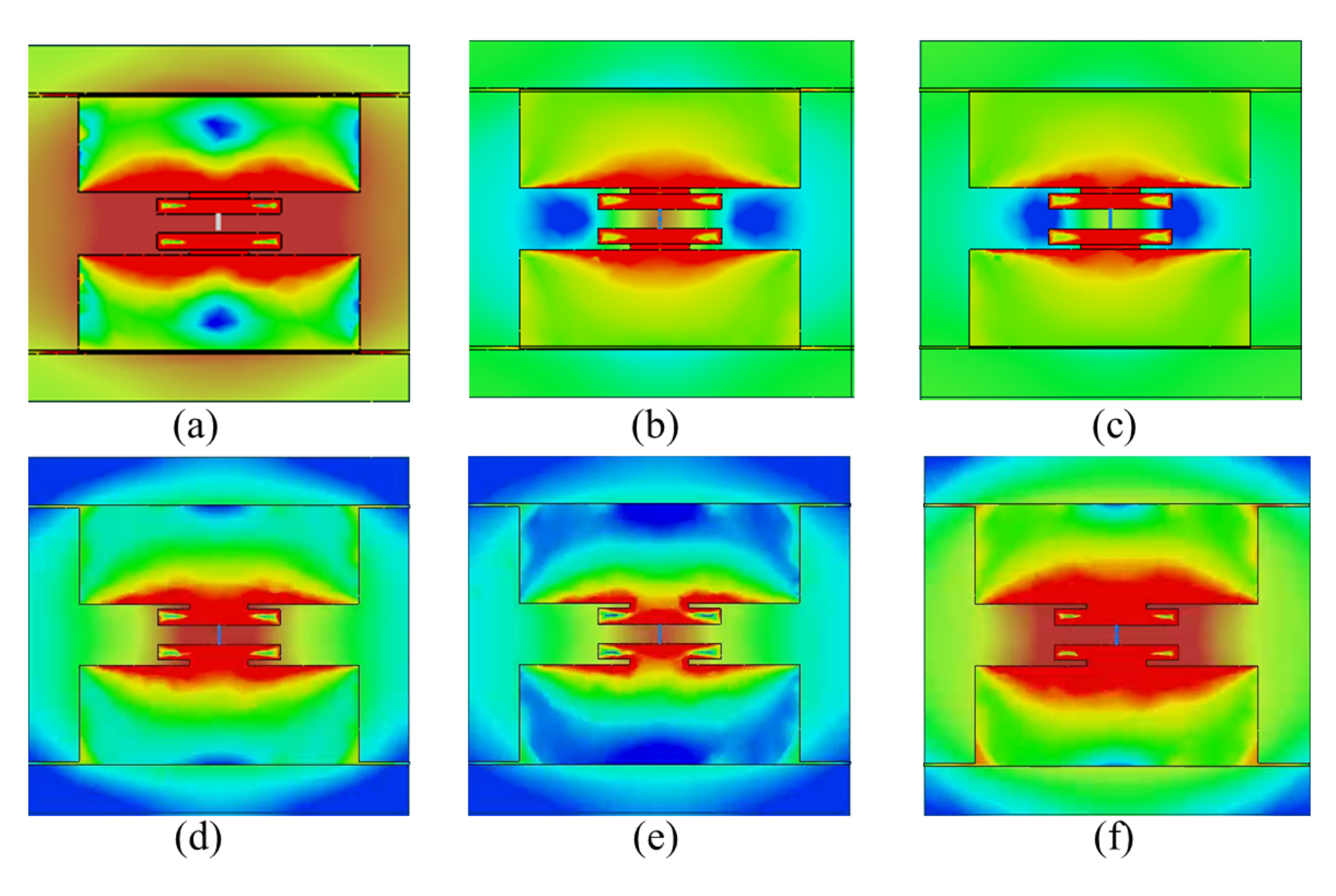}
  %  \vspace{-15pt}
  \caption{Different current distributions. (a), (b), and (c) correspond to different tunable device parameters at the same moment, while (e), (f), and (g) represent those at different moments, namely 0, $T/8$, and $T/4$, with the same tunable device parameters. Here, $T$ is the operating frequency period.  }
  \vspace{0.1cm}
\end{figure}

\subsection {Current distribution image segmentation using Otsu threshold method}

Since the pixel values of the $\boldsymbol{D}$ matrix (obtained as described in Section II-A) range from 0 to 255 and thus hold a substantial amount of information, while in the physical domain of the RIS element, only two configurations are possible, namely the existence or nonexistence of a metal patch. Therefore, it becomes essential to perform image segmentation in $\boldsymbol{D}$ so that each pixel within the image can be assigned a binary value of '0' or '1'.his work, we use threshold segmentation to classify the matrix pixel values $\boldsymbol{D}$ into two states: “0” or “1”, where a value of “1” indicates the presence of a metal patch, and conversely a value of “0” indicates its absence. We define $\boldsymbol{P}$ as the matrix after performing the aforementioned segmentation. The segmentation process of converting the image information into binary information for the pixel $P(i,j)$ in the $i$-th row and the $j$-th column is defined by

According to Eq. (3), a larger value of $O_{t} ^{*}$ implies fewer pixels with a value of “1”, indicating a smaller area of the top patch and a higher degree of miniaturization of the RIS element. This also increases the likelihood of the RIS's frequency shifting towards higher bands, resulting in degraded performance. Conversely, a smaller value of $O_{t} ^{*}$  corresponds to a lower degree of miniaturization of the RIS element but a higher probability of better performance. 
Therefore, to balance the trade-off between the performance of the RIS element and miniaturization, we need to select an appropriate value of $O_{t} ^{*}$.

\begin{equation}
P(i, j)=\left\{\begin{array}{ll}
1, & \text { if } \quad G_{r}(i, j) \geq O_{t}^{*} \\
0, & \text { otherwise }
\end{array}\right.
\end{equation}

Traditional threshold segmentation involves selecting an appropriate threshold through multiple experiments, which results in low stability of segmentation and incurs significant time costs \cite{r14}. In this work, we employ the Otsu threshold segmentation method to determine $O_{t} ^{*}$ , an automatic technique that maximizes the variance between classes between the target and background. This maximization allows the Otsu algorithm to find an optimal threshold distinguishing the target (high current distribution) from the background (low current distribution) in image $\boldsymbol{D}$ \cite{r41}.  The inter-class variance is defined by
\begin{equation}
\frac{\delta _{B}^{2}}{\delta _{W}^{2}}  =(\mu _{1}-\mu _{2})^{2} /(\delta _{1}^{2}+\delta _{2}^{2})
\end{equation}
\begin{equation}
\delta _{B}^{2}+\delta _{W}^{2}=\delta ^{2}
\end{equation}
where $\delta _{B}^{2}$ and $\delta _{W}^{2}$ are the inter-class variance and the intra-class variance, respectively. $\mu _{1}$ and $\mu _{2}$ are means of two classes, respectively. $\delta _{1}$ and $\delta _{2}$ are the variances of two classes, respectively. $\delta$ is the variance of an image, which is a constant value.

Then, the inter-class variance is maximized using Eq(7). By using the Otsu thresholdding method to select $O^*_t$, we naturally balance the trade-off between miniaturization and performance of RlS elements.
\begin{equation}
O_{t} ^{*} =\underset{O_{t}}{argmax} \quad \delta _{B}^{2}(O_{t})
\end{equation}
where $O_{t} $ is the threshold value, and $O_{t} ^{*}$ is the optimal threshold value.

\subsection {Converting the image matrix into RIS elements}

To convert a 2D image into an RIS structure, the fundamental step is to transform the information from the image domain into the physical structure of operational metal patches. In this process, each pixel $P(i,j)$ of the matrix $\boldsymbol{P}$ representing the image has two possible values: “0” and “1”. Specifically, we need to map those elements of $P(i,j)$ that are equal to 1 to the physical structure of RIS, which is characterized by a metallic structure. However, it should be noted that considering the large number of pixels in $\boldsymbol{P}$, directly mapping the pixels with a value of 1 into metal patches would be a time-consuming task. The size of matrix \textit{\textbf{P}} is related to the clarity of the current distribution image. To ensure the clarity of the image, we typically utilize images with a high resolution, which results in a huge size of \textit{\textbf{P}}. In this work, the size of matrix $\boldsymbol{P}$ is $551 \times 586$, which means that there are 322886 pixels. It will incur a significantly high time cost if we directly map the pixels of $\boldsymbol{P}$  to pixel patches of the RIS element.  

To address the above technical challenge, in this study, we sparsify the matrix $\boldsymbol{P}$  through the down-sampling method \cite{r42}, aiming to reduce the time cost of mapping the image into metal patches. If the size of $\boldsymbol{P}$ is $I \times J$, then the size of the sparse matrix $\boldsymbol{P}_t$ is $[I/\gamma_x] \times [J/\gamma_y]$, where $\gamma_x$ and $\gamma_y$ are the row and column sparsity factors of $\boldsymbol{P}$, respectively, and both of them are positive integers. Here, the operation [$\bullet $] is used to obtain the integer closest to $\bullet$. Larger values of $\gamma_x$ and $\gamma_y$  imply a smaller size for matrix $\boldsymbol{P}_t$, resulting in a reduced number of pixels that need to be mapped into pixel patches. Consequently, the time cost for modeling decreases, but the level of detail in the depiction of the current distribution image diminishes. 

Based on matrix $\boldsymbol{P}_t$, preprocessing operations including denoising are performed to enhance image quality. Then, the feeding structure of the RIS element is considered. It is crucial to ascertain whether this structure is connected, as it directly impacts the functionality of the overall system. In the event that it is found to be disconnected, a corrective measure is taken: the pixel values of the matrix corresponding to the feeding position are set to 1. Through this comprehensive process, we successfully obtain the modified matrix  $\boldsymbol{P}_a$. To visually illustrate the transformation, Fig. 4(a) presents an image that corresponds to  $\boldsymbol{P}_t$ matrix. Meanwhile, Fig. 4(b) showcases the image after undergoing the aforementioned operations, including denoising and the refinement of the feeding structure that corresponds to $\boldsymbol{P}_a$ matrix.  

% ==== FIG4
\begin{figure}%[!htbp]
\centering
  \includegraphics[scale=0.54]{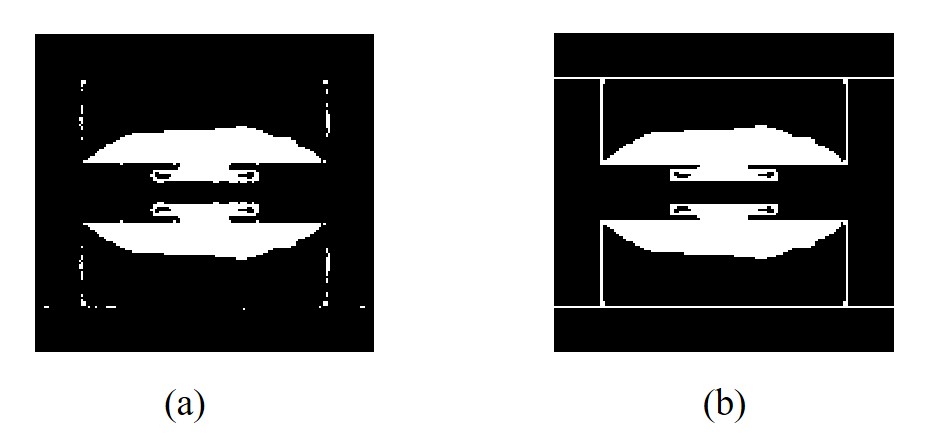}
  %  \vspace{-15pt}
  \caption{(a) Image $\boldsymbol{P}_t$ after sparse operation, and (b) Image $\boldsymbol{P}_a$ after denoising and the addition of the feeding structure.  }
\end{figure}

For $\boldsymbol{P}_a$, we convert a pixel value of 1 into a metal patch in the RIS structure, where the length and width of the metal patch are set to $l_x$ and $l_y$, which are defined by 
\begin{equation}
l_x=\gamma_x \cdot L_d/I
\end{equation}
and 
\begin{equation}
l_y=\gamma_y \cdot W_d/J
\end{equation}
where $L_d$ and $W_d$ are the maximum length and width of the RIS element, respectively. $I$ and $J$ denote the number of rows and columns in the matrix $\boldsymbol{P}_a$, respectively.

If the pixel value of the element in the $P_t$ matrix at the $i$-th row and $j$-th column is 1, we can obtain the metal patch that is mapped onto the physical structure of the RIS element. The starting point $l_x^s(i)$ and the ending point $l_x^e(i)$ of the metal patch along the $x$-axis are as follows:  
\begin{equation}
l_x^s(i)=(i-1)\cdot l_x
\end{equation}
\begin{equation}
l_x^e(i)=i\cdot l_x
\end{equation}

And the starting point $l_y^s(i)$ and the ending point $l_y^e(i)$ of the metal patch along the $y$-axis are as follows:  
\begin{equation}
l_y^s(j)=(j-1)\cdot l_y
\end{equation}
\begin{equation}
l_y^e(j)=j\cdot l_y
\end{equation}

Then, Matlab software is used to call CST software for modeling a rectangular metal patch starting from ($l_x^s(i)$, $l_x^e(i)$) and an ending point of  ($l_y^s(j)$, $l_y^e(j)$). 

The specific algorithm and linear mapping process are shown in Algorithm 1. Utilizing Algorithm 1, it is possible to convert the two-dimensional representation of the current distribution across the RIS elements into a novel topological configuration of these elements.

\begin{algorithm}
\caption{Method for converting 2D image into the RIS element}
\begin{algorithmic}[1]
\REQUIRE The matrix $\boldsymbol{P}_a$ of the image, the dielectric constant, loss, and thickness of the dielectric plate, the minimum unit length $l_x$ and width $l_y$ of topology design, and the number of rows $[I/\gamma_x]$ and columns $[J/\gamma_y]$ in the matrix $\boldsymbol{P}_a$.
\ENSURE New topology structure of RIS elements.

\STATE Initialization: Modeling the ground and the substrate on CST software.
\STATE $i=0$, and $j=0$.

\WHILE{$j<[J/\gamma_y]$}
    \STATE $j=j+1$.
    \WHILE{$i<[I/\gamma_x]$}
        \STATE $i=i+1$.
        \IF{$P(i, j)=1$}
            \STATE Determine the starting point $l_x^s(i)$ of the $x$-axis using Equation (10).
            \STATE Determine the ending point $l_x^e(i)$ of the $x$-axis using Equation (11).
            \STATE Determine the starting point of the $y$-axis using Equation (12).
            \STATE Determine the ending point of the $y$-axis using Equation (13).
            \STATE Use Matlab software to call CST software for modeling a rectangular metal patch starting from ($l_x^s(i)$, $l_x^e(i)$) and an ending point of ($l_y^s(j)$, $l_y^e(j)$).
        \ELSE
            \STATE Do not perform any operation.
        \ENDIF
    \ENDWHILE
\ENDWHILE
\end{algorithmic}
\end{algorithm}

\subsection {Optimization for different working states in RIS elements and comprising the RIS array}

Once the topology of the top metal patch is determined, we add the substrate and ground to outline the configuration of the RIS element. Furthermore, to attain various operational states, it is necessary to obtain the parameters of equivalent circuit for the tunable device $\bm{x}$.
Let us denote the $i$-th EM amplitude and phase corresponding to operational state of the RIS element at frequency $w$ as $A(S_{11}(\bm{x},\omega))$ and $P(S_{11}(\bm{x}),\omega)$, respectively. Here $S_{11}(\bm{x},\omega)$  is the reflection coefficient at frequency $w$ corresponding to the design variable $\bm{x}$. And the expected amplitude and phase of the corresponding state of the RIS element frequency $w$ are $A^*(S^i_{11}(\omega))$ and $P^*(S^i_{11})$ for $i$th working state, $i$=0, ..., 7.  respectively. The optimization of the corresponding parameters can be expressed as follows:
\begin{equation}
\begin{aligned}
\bm{x}_{i}^{*}=&\underset{\bm{x}}{argmin}\quad {\textstyle \sum_{i=1}^{K}} (  W_1\cdot ||P(S_{11}(\bm{x},\omega^i))-P^*(S^i_{11}(\omega^i))||^2\\
&+W_2\cdot max(|A(S_{11}(\bm{x},\omega^i))|_{dB},|A^*(S^i_{11}(\omega^i))|_{dB}))\\
\end{aligned}
\end{equation}
where $\bm{x}$ is the design variables of the tunable device for the RIS element. $\bm{x}^{*}$ is the optimal parameters of the tunable device for the RIS element. $||\bullet ||^2$ is the 2-norm function of $\bullet $. $|\bullet |_{dB}$ is  the dB form of amplitude of the reflection coefficient. $max(\bullet , \bigtriangleup )$ is the maximum value between $ \bullet $ and $\bigtriangleup$. $w_1$ and $w_2$ are the proportions of phase and amplitude in the objective function, with a variation range between 0-1. In this work, $W_1$ = $W_2$ =0.5. $\omega^i$ is the $i$-th considered frequency point, $i$=1, ..., $K$, and $K$ is the number of considered frequency points.

Due to the limited dimensions of the equivalent parameters of tunable devices, we use the quasi-Newton method\cite{rr15} to obtain the optimal equivalent parameters of tunable devices. Since the quasi-Newton method requires an initial value to be given, we set the initial value to be determined by the equal division of tunable devices, as shown in the following equation.

\begin{equation}
\bm{x}_{(i)}=\bm{x}_{min}+\frac{(\bm{x}_{max}-\bm{x}_{max})}{2^N}\cdot(i-1)
\end{equation}

The quasi-Newton method has good local optimization ability and the ability to quickly converge to the local extremum. Furthermore, it does not require the calculation of the second derivative and its inverse. This algorithm can be described as:

\begin{equation}
\bm{x}_i^{t+1}=\bm{x}_i^{t}+\bm{H}^{t+1}\cdot \Delta \bm{g}^t
\end{equation}
where $\bm{x}_i^{t+1}$ and $\bm{x}_i^{t}$ are the optimal solutions for the ($t$+1) and $t$ generations, respectively.
$\bm{H}^{t+1}$ is the Hessian matrix of the ($t$+1) generations.
$\Delta \bm{g}^t$ is the change in input $\bm{x}_i$ for the $t$-th generation.

Subsequently, using the RIS elements in various states, we derive the RIS array. By applying the generalized Sennels' theorem, we can facilitate arbitrary beamforming encoding, thereby allowing for control over the orientation of the transmitted wave beams.

\section{Application and Results} 
This section presents an evaluation result of the performance of the proposed method through the design and implementation of 3-bit phase-modulated RISs. The implementation of the proposed method is carried out using the Matlab platform and CST software on a computing system that features an AMD Ryzen 5500 CPU and 32 GB of RAM.

\subsection { Simulation results}

In this work, a 3-bit RIS element is used as the initial structure (shown in Fig. 5 (a)),  and the operating frequency is at 2.6 GHz. The initial structure is divided into four layers: a layer with tunable components, a layer of passive metal patch, another layer of substrate (an F4B substrate, \textit{ε\textsubscript{r} }= 2.65, tan\textit{\delta} = 0.001, thickness is 2 mm), and a ground plane. There are two metallic girded patches bridged by a varactor diode (SMV-2019-079LF) on the top of the RIS element. The equivalent circuit of the varactor is a series RLC circuit, in which \textit{R} = 0.3 $\Omega$, \textit{L} = 0.7 nH, and \textit{C} is the equivalent capacitance as a function of bias voltage. The design variable of the tunable device is the equivalent capacitance value $\bm{x}$= $C$, ranging from 0.6 pF to 2.6 pF. Specifically, 2.6 pF represents the capacitance value when the varactor is in an open-circuited state, while 0.6 pF corresponds to the capacitance value at the maximum operating voltage applied by the user. 
Based on the initial structure, we use the proposed method to achieve a new topology structure, as depicted in Fig. 5(b). Within this method, specific parameters are set: $\gamma_x$ = $\gamma_y$ = 4, while $O_{t} ^{*}$ is set to 203. 

% =======
% FIG. 5
% =======
\begin{figure}
\centering
\subfloat[]{\includegraphics[width=2.4in]{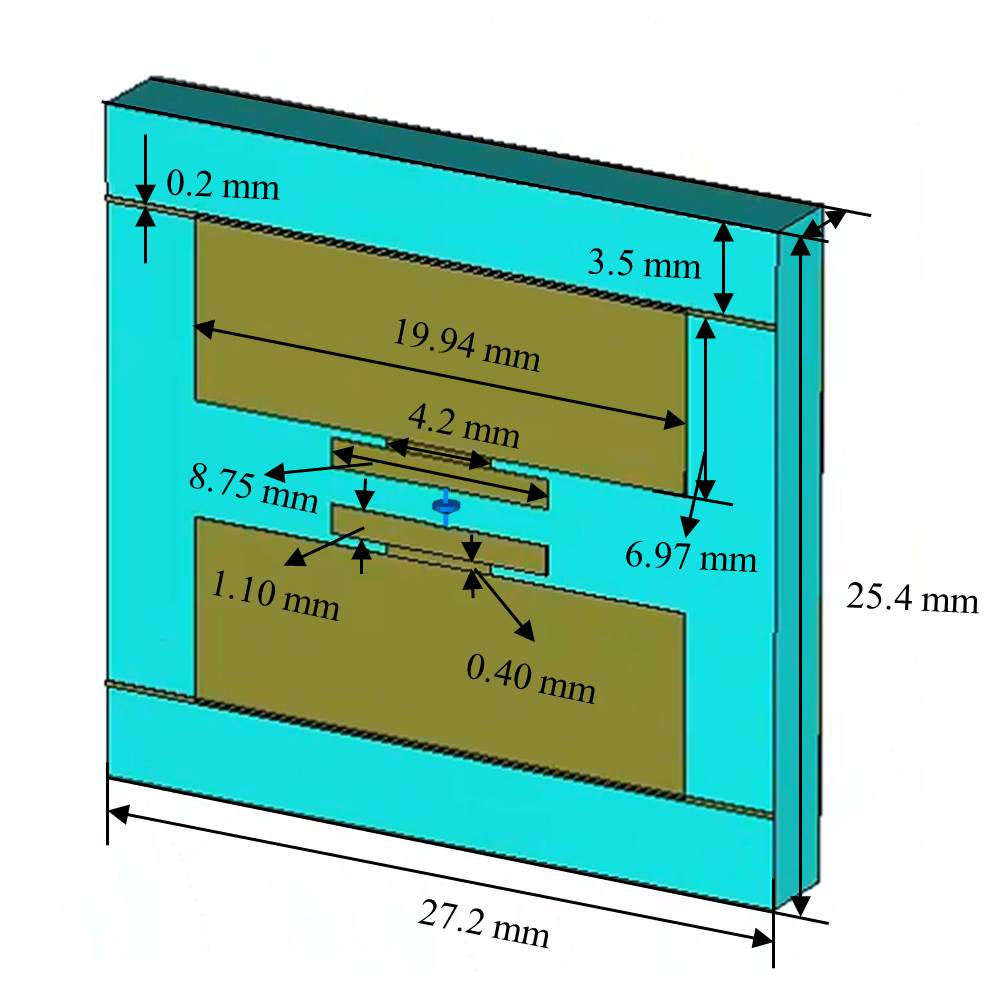}}

\subfloat[]{\includegraphics[width=2.4in]{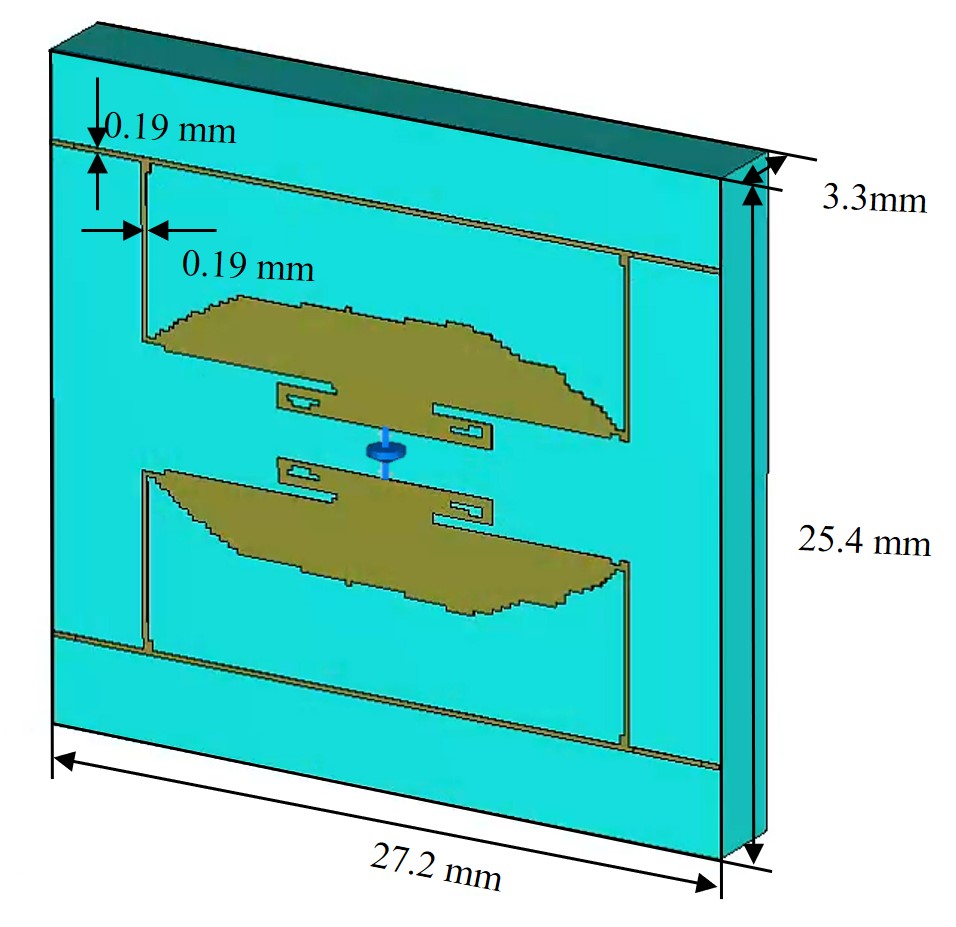}}
\caption{ (a) Initial structure of RIS elements, and (b) New structure of RIS elements.}
\end{figure}

The reflection coefficients of the eight working states of the initial structure are presented in Fig. 6. This initial structure exhibits 8 phase states, with the amplitude of the reflection coefficient exceeding -1.9 dB at the working frequency. Detailed information regarding the phases, amplitudes, and corresponding capacitance values of these eight working states is listed in Table I.
Subsequently, turning our attention to the optimal structure, the reflection coefficients of its eight working states are depicted in Fig. 7. Similar to the initial structure, it likewise possesses eight working states, wherein the amplitude of the reflection coefficient at the working frequency point is greater than -1.4 dB. The relevant data for the phases, amplitudes, and corresponding capacitance values of these eight working states are enumerated in Table II. It can be seen that the three-bit phase modulation feature is well positioned in the new topology RIS. In terms of miniaturization, the new structure we propose has a top metal patch area that accounts for only 15 \% of the entire area, compared to 65 \% occupied by the original structure. This demonstrates that our new structure has achieved significant miniaturization.

Based on the aforementioned content, we can obtain the RIS element under eight different states. Subsequently, utilizing the generalized Sennels' theorem, we can achieve arbitrary beamforming encoding, enabling precise control over the direction of the transmitted wave beams.

% =======
% FIG. 6
% =======
\begin{figure}
\centering
\subfloat[]{\includegraphics[width=3in]{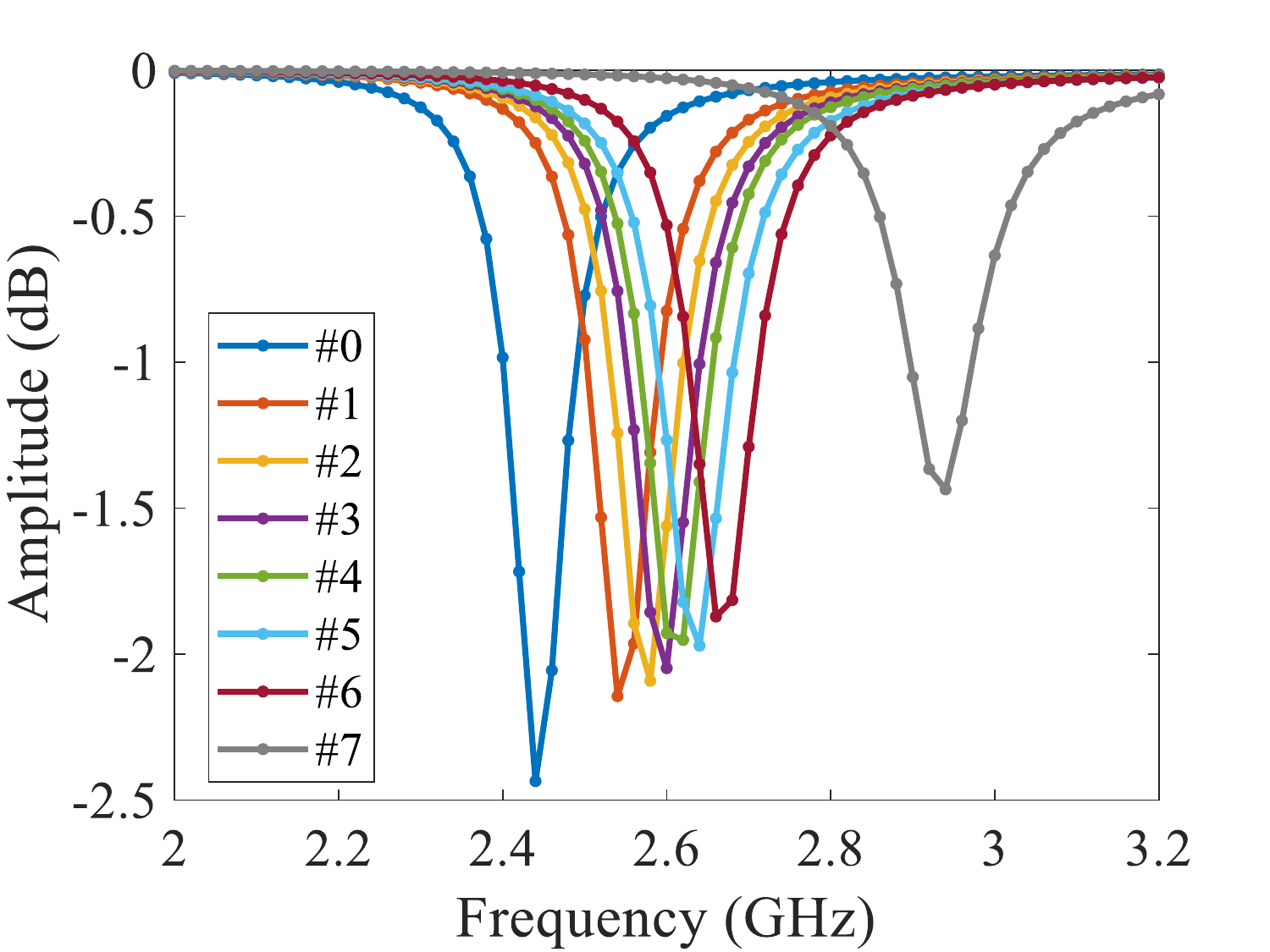}}

\subfloat[]{\includegraphics[width=3in]{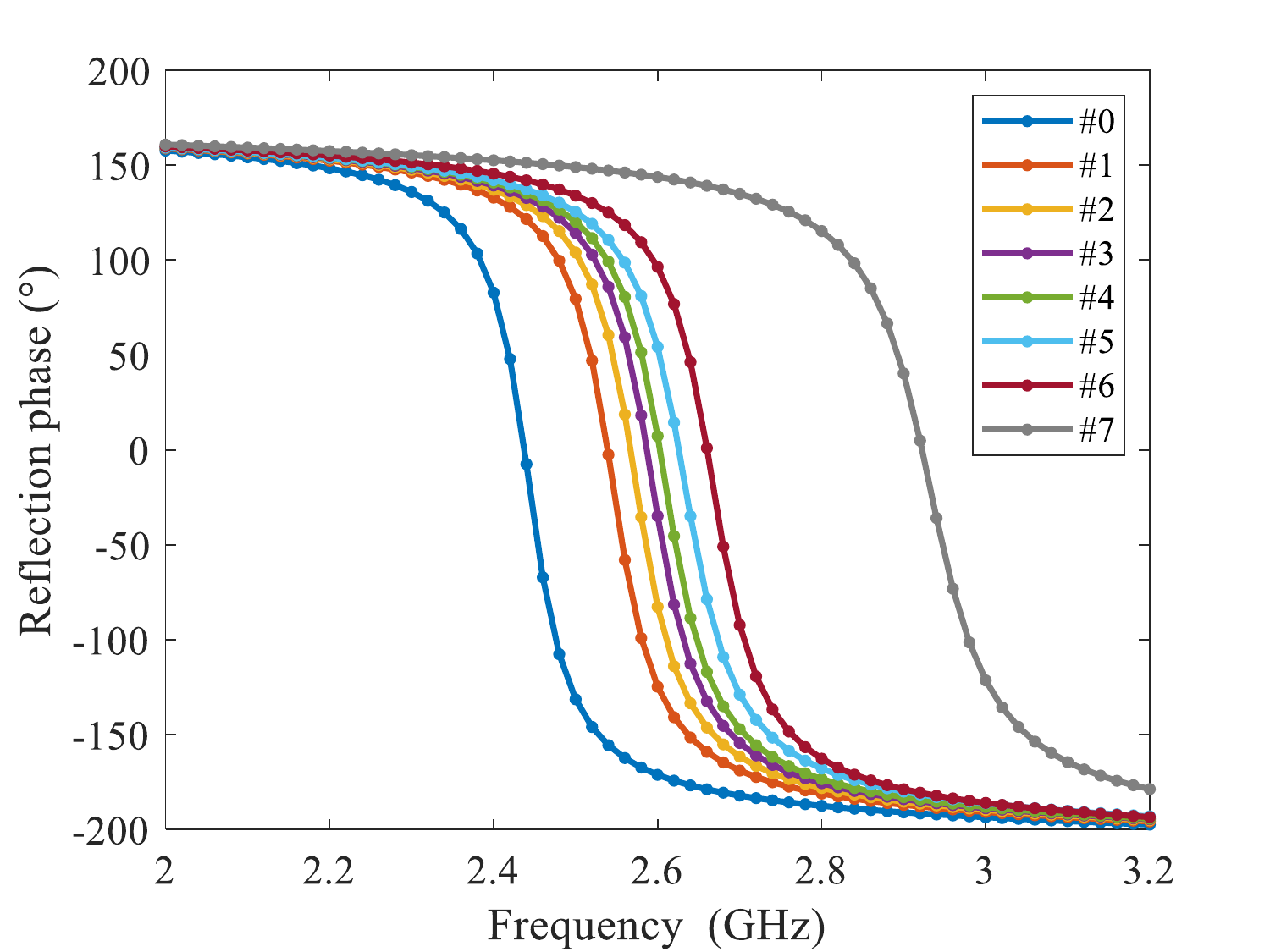}}
\caption{ The reflection coefficient using EM simulation of RIS elements for the initial structure,  (a) amplitude, and (b) phase.}
\end{figure}

\begin{table}[!ht]
    \centering
    \renewcommand\arraystretch{1.5}
    \caption{Parameters of working states for the initial RIS element}
    \scalebox{1.05}{
    \begin{tabular}{clllcc}
    \hline
        Working states&    $R$&$L$&$C^*$&Amplitude (dB)& Phase\\ \hline
        0\#&    0.3&0.7&2.60&-0.07& 0°\\
 1\#&   0.3&0.7&2.02& -0.21&46°\\
 2\#&   0.3&0.7&1.93& -0.32&89°\\
 3\#&   0.3&0.7&1.86& -0.45&136°\\
 4\#&   0.3&0.7&1.79& -0.60&178°\\
 5\#&   0.3&0.7&1.91& -1.04&225°\\
 6\#&   0.3&0.7&1.49& -1.81&268°\\ 
        7\#&    0.3&0.7&0.98&-0.05& 315°\\ \hline
    \end{tabular}
    }

\end{table}

\begin{table}[!ht]
    \centering
    \renewcommand\arraystretch{1.5}
    \caption{Parameters of working states for the optimal RIS element}
    \scalebox{1.05}{
    \begin{tabular}{clllcc}
    \hline
        Working states&    $R$&$L$&$C^*$&Amplitude (dB)& Phase\\ \hline
        0\#&    0.3&0.7&2.60&-0.02& 0\\
 1\#&   0.3&0.7&1.08& -0.33&46\\
 2\#&   0.3&0.7&0.99& -0.84&91\\
 3\#&   0.3&0.7&0.98& -1.27&139\\
 4\#&   0.3&0.7&0.95& -1.33&180\\
 5\#&   0.3&0.7&0.89& -1.00&229\\
 6\#&   0.3&0.7&0.84& -0.54&268\\ 
        7\#&    0.3&0.7&0.72&-0.10& 315\\ \hline
    \end{tabular}
    }

\end{table}

% =======
% FIG. 7
% =======
\begin{figure}
\centering
\subfloat[]{\includegraphics[width=3in]{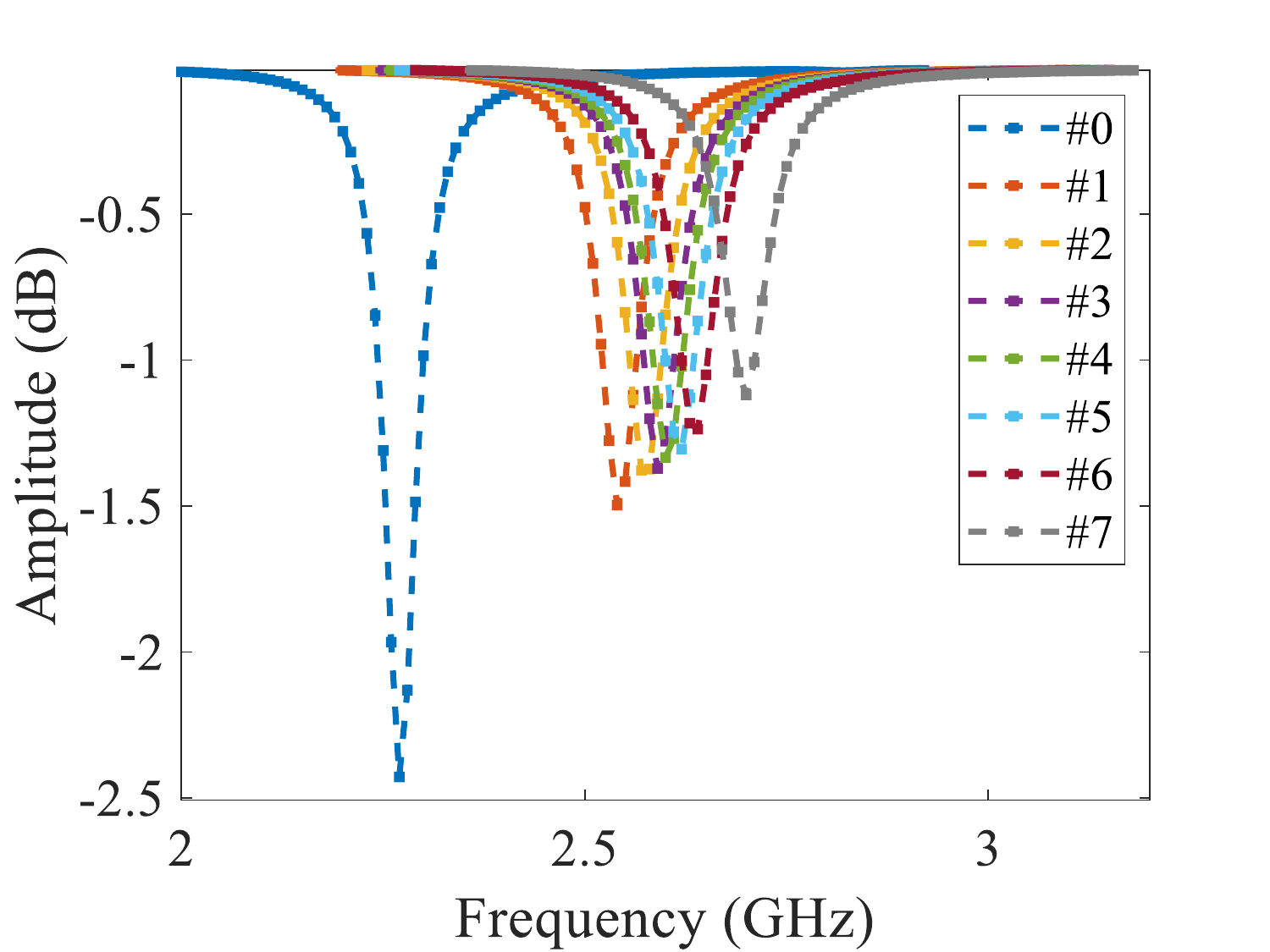}}

\subfloat[]{\includegraphics[width=3in]{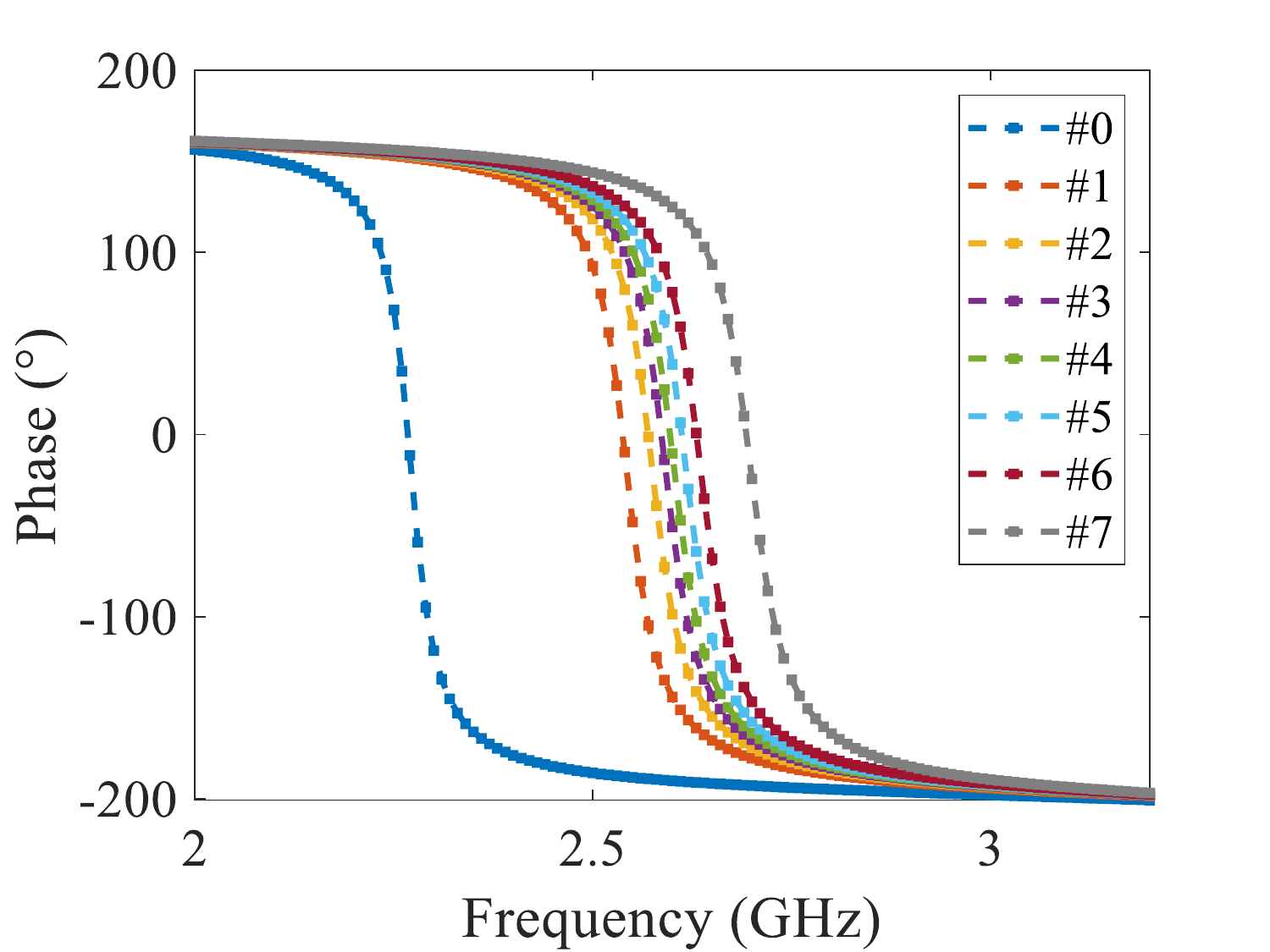}}
\caption{ The reflection coefficient using EM simulation of RIS elements for the optimal structure,  (a) amplitude, and (b) phase.}
\end{figure}

\subsection{Measurement results}

% ==== FIG 8
\begin{figure}[htbp]
% \begin{figure}[htbp]
\centering
\includegraphics[scale=0.11
]{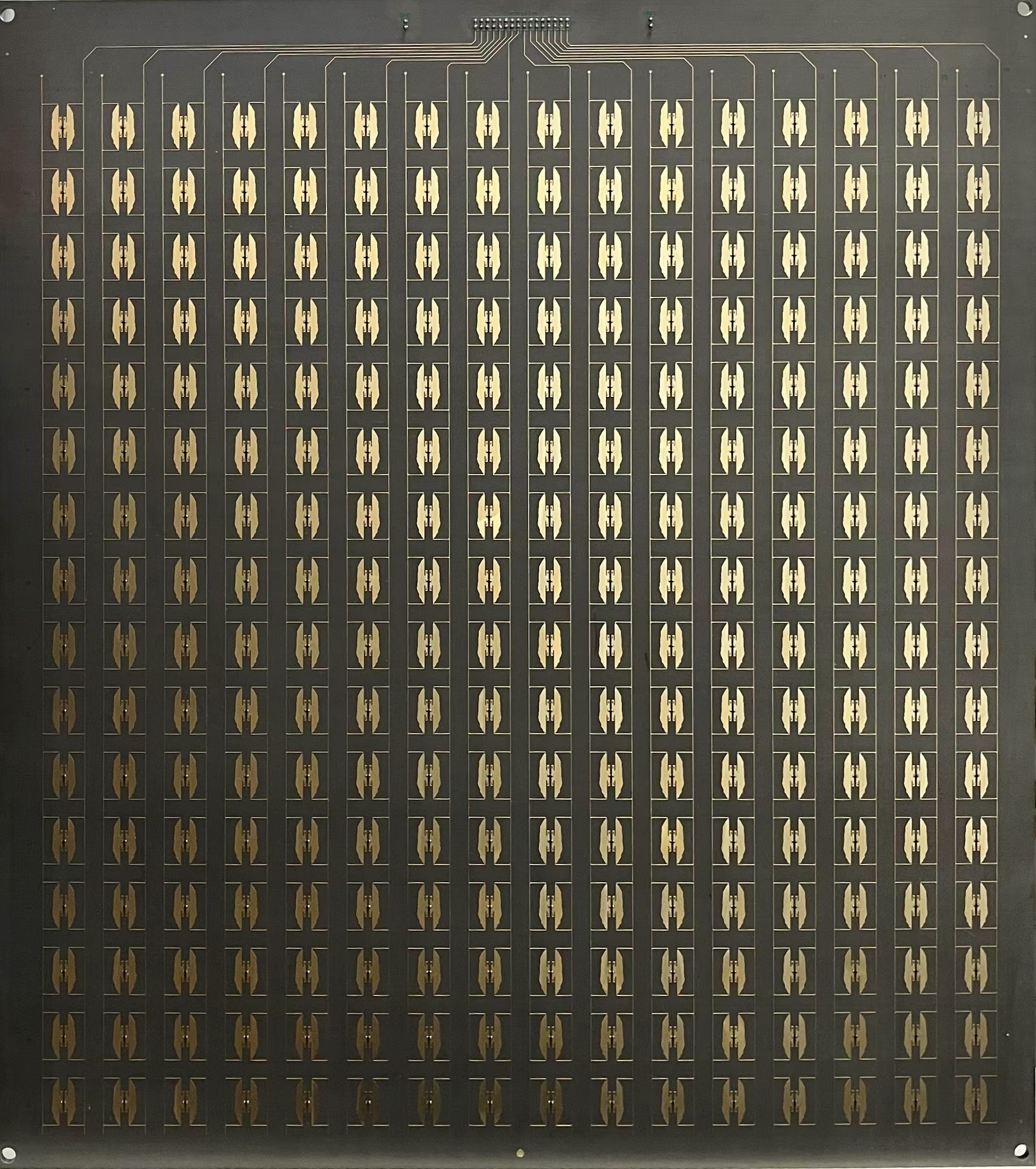}
\caption{Photograph of the fabricated three-bit RIS with 16×16 elements.}
\end{figure}

A 3-bit RIS prototype with a configuration of 16 $\times$16  elements has been fabricated. The image of the manufactured prototype is displayed in Fig. 8. The overall dimensions of the prototype measure 407 mm $\times$450 mm. A total of 256 varactor diodes are affixed to the RIS sample. The varactors within the same column share a common bias voltage, enabling the working state of each column to be modified simultaneously.

Initially, we perform tests on the reflection coefficients of the RIS. In the frequency range from 2 GHz to 3.2 GHz, we carry out measurements of the reflection coefficients of the RIS under diverse bias voltages. These measurements are executed in a microwave anechoic chamber. A vector network analyzer (VNA, Agilent N5245A), a DC voltage source, and a linearly polarized horn antenna are utilized in the measurement process. The horn antenna connected to the VNA is positioned on the normal axis of the RIS. The DC voltage source steadily supplies the same biasing voltage to all diodes of each column on the RIS. A metal sheet of identical size as the RIS is measured using the same experimental setup to normalize the reflection amplitude and phase. The test outcomes of the reflection coefficients are presented in Fig. 9. It can be observed that the RIS possesses eight phase states at 2.75 GHz, namely “0°”, “46°”, “92°”, “138°”, “176°”, “215°”, “255°”, and “305°”. The magnitude of the reflection coefficients for all eight states is less than 3.5 dB. When compared with the simulation results, the test result reveals a frequency offset of 0.15 GHz. During the fabrication of the RIS, numerous challenges may emerge. These include alignment inconsistencies, the accumulation of errors during the processing of multilayer samples, and parasitic capacitance that occurs during the welding of adjustable devices. Such issues can lead to discrepancies between the actual structure and the corresponding simulation model, thereby having an impact on the frequency characteristics of the RIS.

% =======
% FIG. 9
% =======
\begin{figure}
\centering
\subfloat[]{\includegraphics[width=3in]{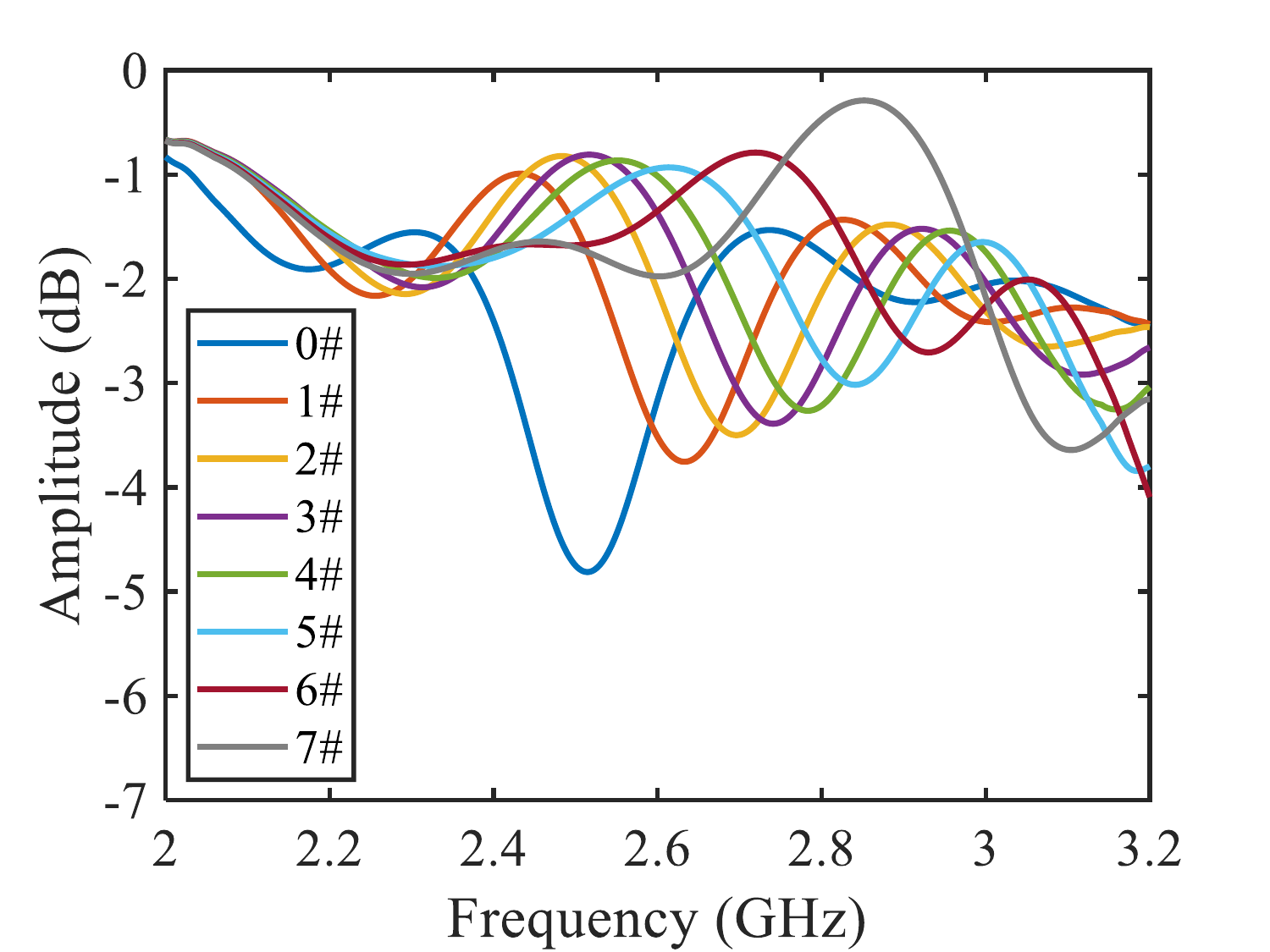}}

\subfloat[]{\includegraphics[width=3in]{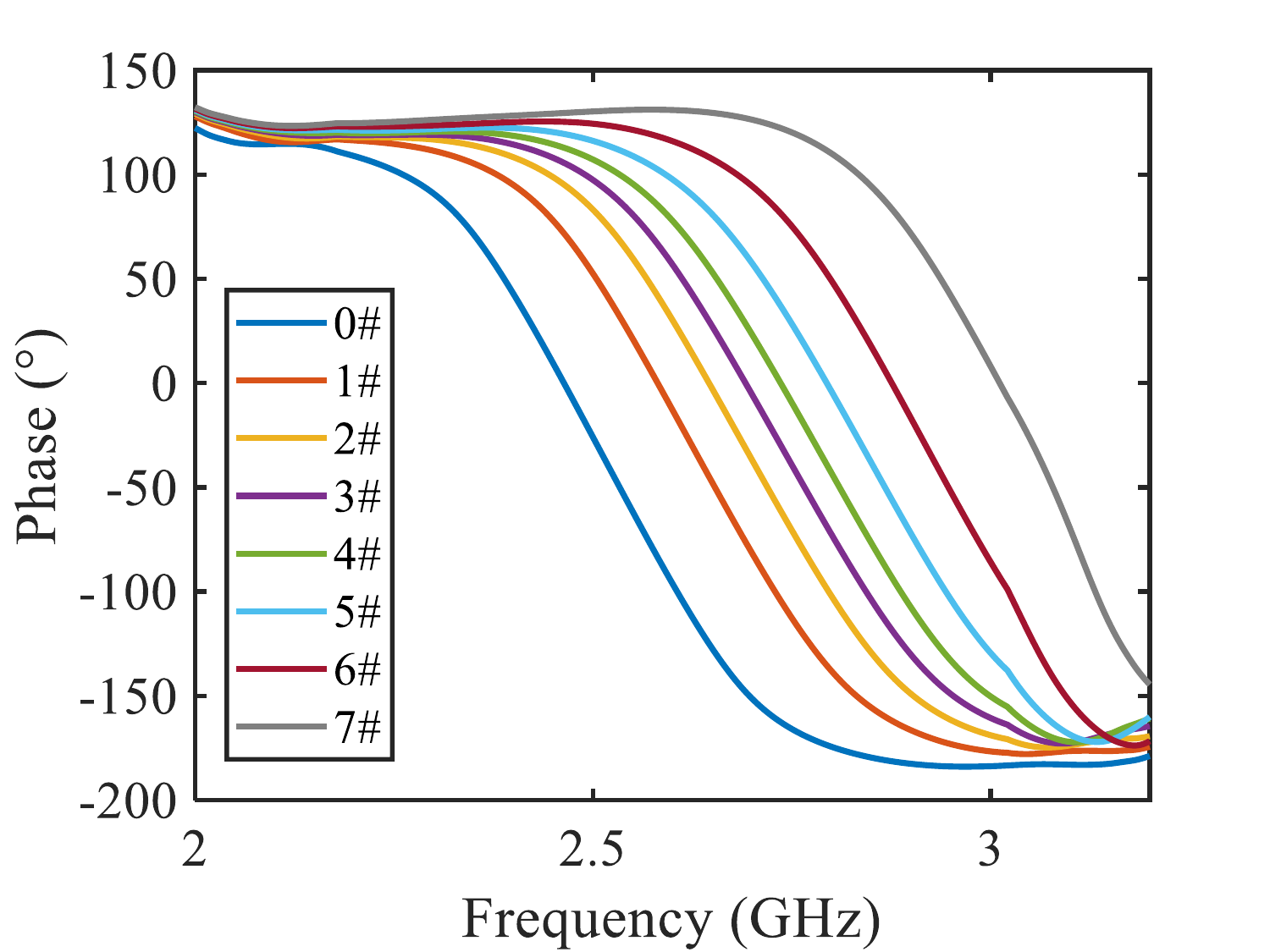}}
\caption{ The measured reflection coefficient of the RIS,  (a) amplitude, and (b) phase.}
\end{figure}
% \vspace{-1.0cm}

Based on the analysis of the test reflection coefficients of the RIS mentioned above, we measured the far-field radiation pattern of the RIS at 2.75 GHz.
The measurement is conducted in a microwave anechoic chamber. A vector network analyzer is used to connect two horn antennas for transmitting and receiving signals. DC voltage sources are employed to provide the bias voltage for the varactors in the RIS array. Each column of the RIS element is controlled by a dedicated DC voltage source. Therefore a total of 16 DC voltage sources are employed in the experiment. The transmitting antenna and the RIS are mounted on a turntable, while the receiving horn antenna is positioned 8 meters away from the RIS. To mitigate undesired reflections from the surrounding environments, time-gating technology is employed in the experiments. 

In the experiment, the scattering pattern is adjusted by modifying the reflection amplitude and phase of each RIS element. The coding state of each column in the RIS array is determined according to the desired beam direction. The measurement setup is depicted in Fig. 10. An external DC voltage source is utilized to control the phase state of each RIS element by supplying a specific biasing voltage to the diode on the element. Based on the generalized Snell's law, at 2.75 GHz, the beam angle under the coding scheme “4432210077655433” is approximately -20°, and the beam angle under the coding scheme “0000222244446666” is around 15°.

We employ the array synthesis method \cite{r30} to calculate the far-field scattering patterns. The scattering far-field patterns from electromagnetic (EM) simulation and measurement are compared under two coding sequences (“4432210077655433” and “0000222244446666”), as illustrated in Fig. 11. Fig. 11(a) displays the far-field patterns with the main lobe at -$20^{\circ}$ at normal incidence. Fig. 11(b) shows the far-field pattern at normal incidence with the main lobe directed at  $15^{\circ}$. The main lobe position and the number of sidelobes obtained from our method, EM simulation, and measurement are generally in agreement. The slight differences in the sidelobes are mainly due to the fact that the mutual couplings among adjacent RIS elements are not fully considered in our array synthesis method. Additionally, another possible reason is the multipath effect between the transmitting and receiving horn antennas in the experiment. Nevertheless, the predicted and measured scattering patterns show reasonable agreement with each other in Fig. 10, which validates the presented method.

% =======
% FIG. 10
% =======
\begin{figure}[htbp]
% \begin{figure}[htbp]
\centering
\includegraphics[scale=0.34]{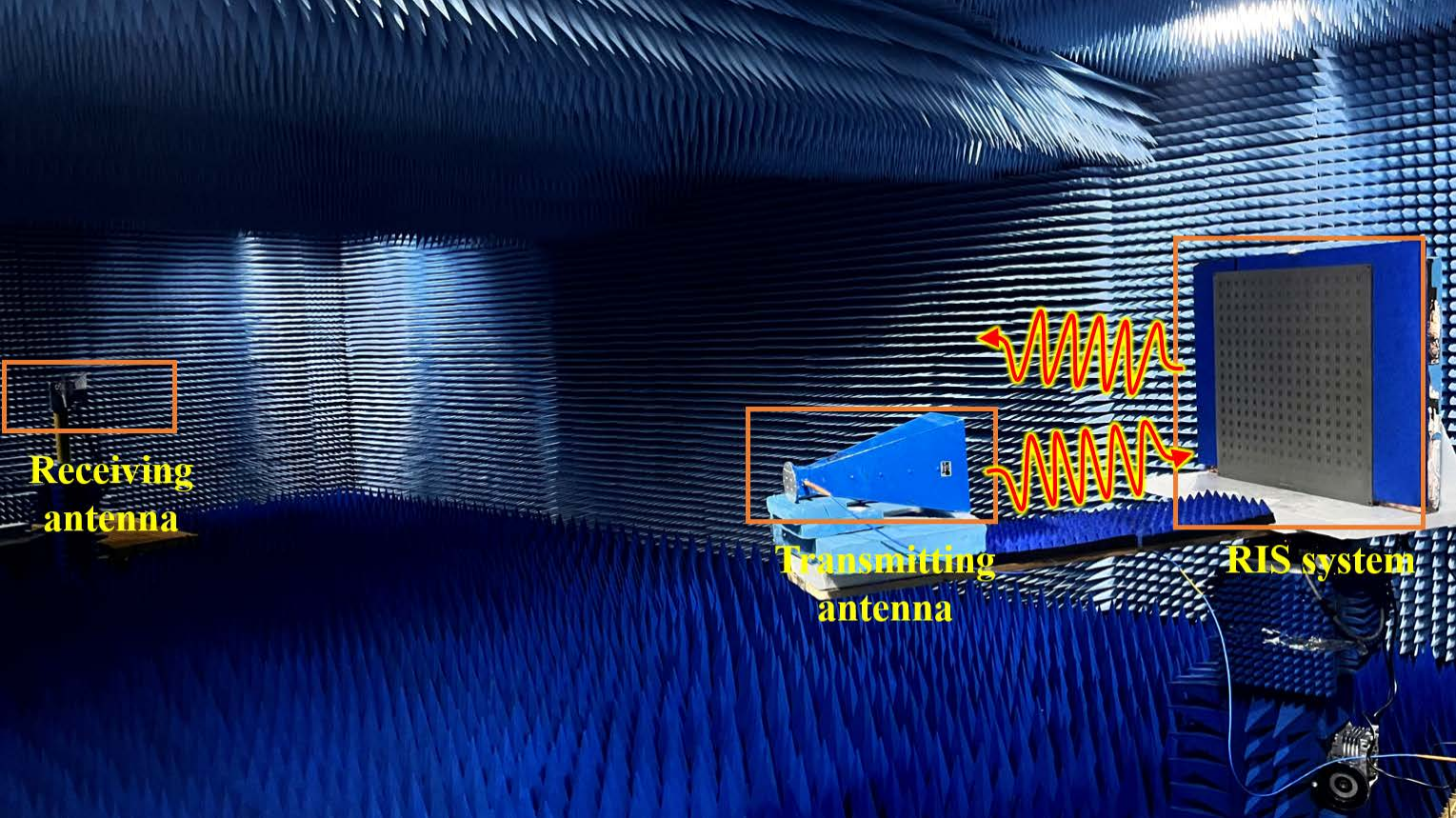}
\caption{Measurement setup to evaluate the far-field scattering pattern of the RIS.}
\vspace{0.1cm}
\end{figure}

% ==== FIG 11
\begin{figure}
\centering
\subfloat[]{\includegraphics[width=3in]{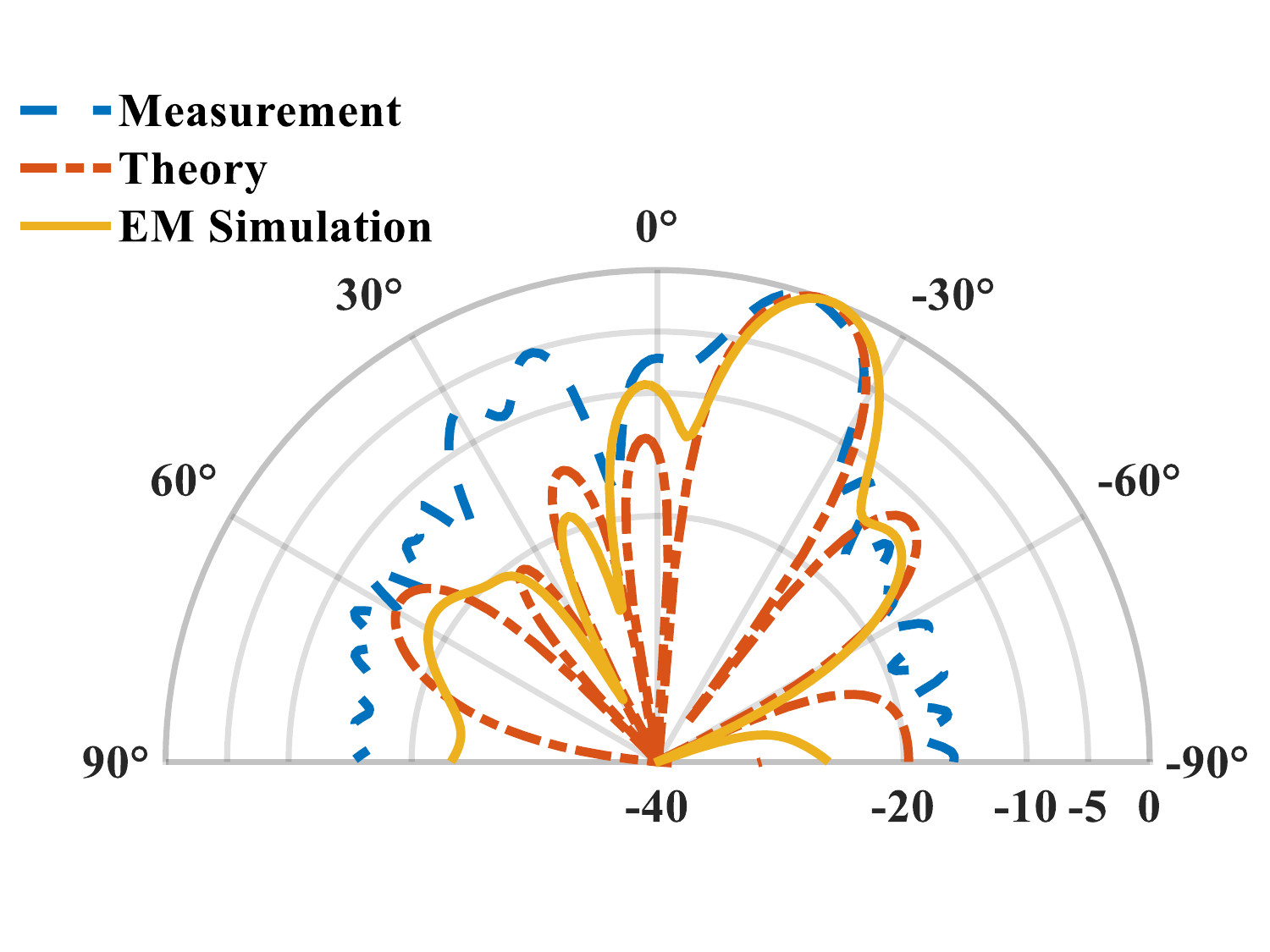}}

\subfloat[]{\includegraphics[width=3in]{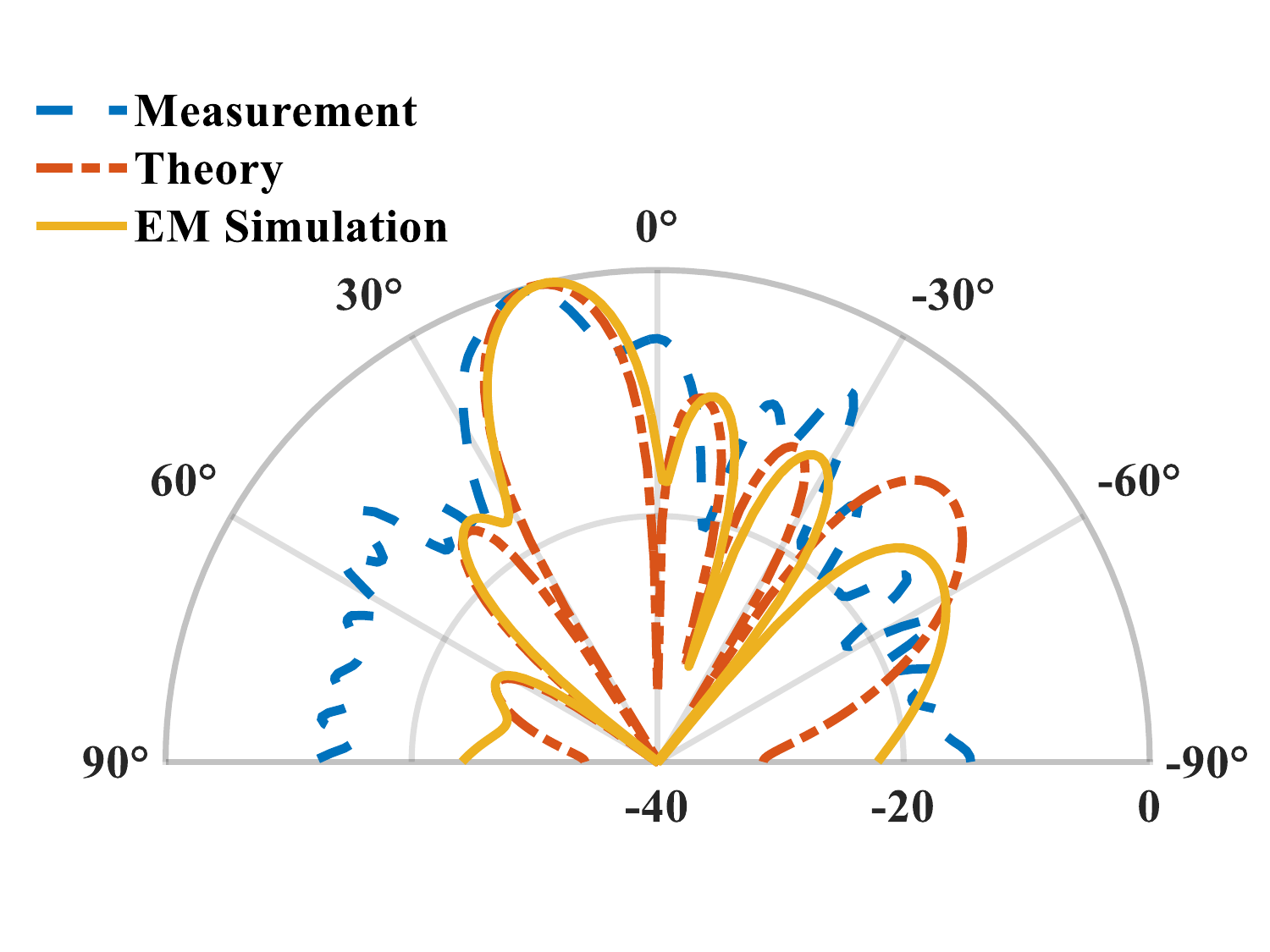}}
\caption{ Comparison of the far-field scattering patterns of the 3-bit RIS with the proposed method, the EM simulation, and the measurement with different phase coding sequences, (a) \ “4432210077655433”, and (b)  “0000222244446666”.}
\end{figure}

\subsection{Comparison with related works}
The proposed topology design method in this work achieves the miniaturization of the top patch of RIS elements, which is very important for RIS systems. For one thing, the miniaturization of the top metal can provide possibilities for the integration design of RIS element \cite{rr17,rr18} including integration of multiple frequency bands for RIS elements, as well as integration of  RIS elements and antenna structures. For another, the miniaturization of the top metal can enhance the transparency \cite{rr19,rr20, rr21} of transparent RISs. To measure the impact of our proposed method on top metal patch miniaturization, we define the miniaturization ratio as the ratio of the area of the top metal patch to the square of the wavelength, as follows:

% ==== FIG 12
\begin{equation}
\varphi =M/\lambda ^2\cdot 100\%
\end{equation}
where $\varphi$ is the miniaturization ratio, $M$ is the area of the top metal patch, and $\lambda$ is the wavelength of the center frequency.

The smaller $\varphi$ , the greater the degree of miniaturization achieved by the RIS structure. Then, when compared to existing works as shown in Table III, it can be seen that, compared to the other 5 works, the miniaturization ratio of the RIS element structure obtained using our proposed topology design method is lower than that of the other 5 structures, indicating the efficiency of our proposed method.

\begin{figure*}[ht]
  \centering
  \includegraphics[scale=0.45]{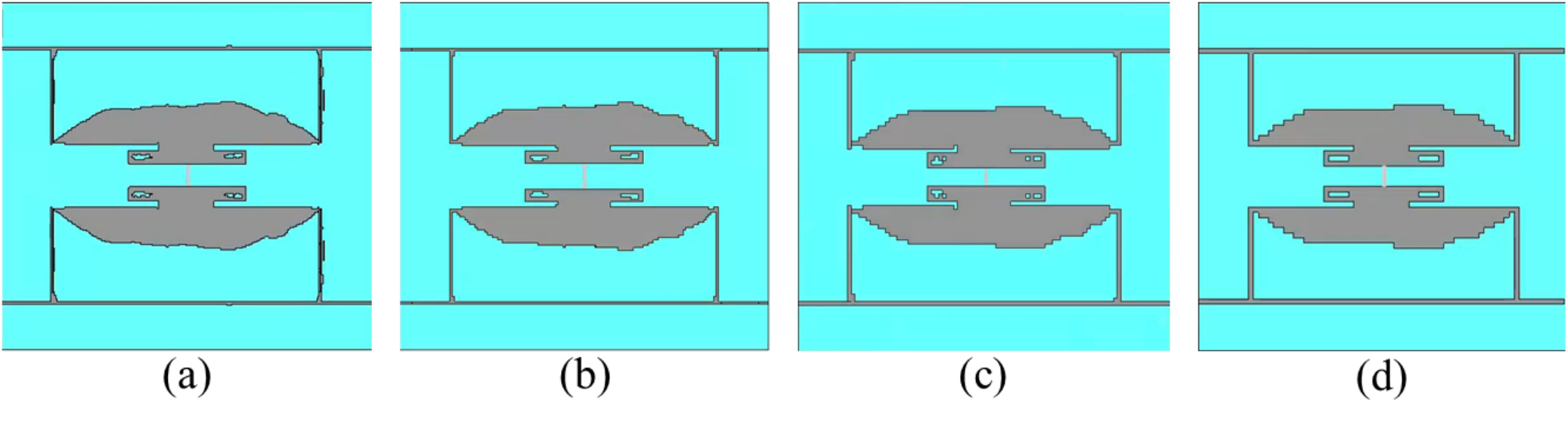}
  %  \vspace{-15pt}
  \caption{RIS element structures with different sparsity factors. (a) $\gamma_x$= $\gamma_y$=2, (b) $\gamma_x$= $\gamma_y$=4, (c) $\gamma_x$= $\gamma_y$=6, and $\gamma_x$= $\gamma_y$=8.} 
  \vspace{0.1cm}
\end{figure*}

\begin{table}[!ht]
    \centering
    \renewcommand\arraystretch{1.5}
    \caption{Performance Comparison of Optimization Methods}
    \scalebox{1.05}{
    \begin{tabular}{cllll}
    \hline
        Ref.&    \makecell[c]{Bit Number} &$\varphi$ &\makecell[c]{Center \\ Frequency (GHz)}&\makecell[c]{Working \\ mode}\\ \hline
        \cite{rr23}&    \makecell[c]{1}&5.68&\makecell[c]{9.4}&reflection\\
 \cite{r9}& \makecell[c]{3}& 3.29&\makecell[c]{3.15}&reflection\\
 \cite{rr9}&   \makecell[c]{4}&1.02&\makecell[c]{2.4}&reflection\\
 \cite{r34}&   \makecell[c]{3}&1.26&\makecell[c]{2.55}&reflection\\
 \cite{rr22}&   \makecell[c]{2}&2.36&\makecell[c]{4.25}&reflection\\ 
        This work&    \makecell[c]{3}&0.86&\makecell[c]{0.75}&reflection\\ \hline
    \end{tabular}
    }

\end{table}

\subsection{Discussion on the influence of the sparsity factor}
To further illustrate the significance of our proposed method, in this work we discuss the influence of the sparsity factor on topology structure for RIS elements. 

In Section II. C, $\gamma_x$ and $\gamma_y$  are  the row and column sparsity factor of $P$, which determine the size of the pixel patch unit. To investigate the impact of $\gamma_x$ and $\gamma_y$ on the topology structure, we model the different structures with different  $\gamma_x$ and $\gamma_y$ values, and calculate the modeling time ( Matlab calling CST to obtain the new topology of the top metal patch) and the size of the pixel metal patch unit, as shown in Table III. 
From this Table, it can be seen that as $\gamma_x$ and $\gamma_y$ increase, the size of the patch unit becomes larger, and the modeling time becomes shorter. In order to further explore the impact of  $\gamma_x$ and $\gamma_y$ factors on the characterization of the topology structure, we generate RIS elements using our proposed topology design method with different $\gamma_x$ and $\gamma_y$ values, as shown in Fig. 12. It can be observed that the smaller the sparsity factor, the clearer the characterization of the RIS element. However, the time cost for modeling increases at this point as shown in Table IV. Therefore, this paper provides an empirical value: when both the length and width of the smallest patch are less than 1/400 of the wavelength, it is recommended to balance the modeling cost and the accuracy of the modeling. Considering that the resolution difference between $\gamma_x$ = $\gamma_y$ = 4 and $\gamma_x$ = $\gamma_y$ = 2 is insignificant, yet $\gamma_x$ = $\gamma_y$ = 4  is more time-consuming in terms of computational cost, the structure with $\gamma_x$ = $\gamma_y$ = 4 is therefore adopted as the designed structure in Sections III.A and III.B.

\begin{table}[!ht]
    \centering
    \renewcommand\arraystretch{1.5}
    \caption{Comparison of Different Sparsity Factors}
    \scalebox{1.05}{
    \begin{tabular}{clllc}
    \hline
        &     $\gamma_x$= $\gamma_y$=2&$\gamma_x$= $\gamma_y$=4&$\gamma_x$= $\gamma_y$=6&$\gamma_x$= $\gamma_y$=8\\ \hline
        Time (s)&    770&67&20&11\\
 $lx$ (mm)&   0.093&0.185&0.278& 0.368\\ 
        $ly$ (mm)&    0.094&0.187&0.280&0.374\\ \hline
    \end{tabular}
    }

\end{table}

\section{Conclusions}

Based on the comprehensive investigation of the current distribution's pivotal role in enhancing the performance of reconfigurable intelligent surface (RIS) elements, this study has successfully developed a novel topology design methodology. By leveraging current distribution analysis and image processing techniques, we have demonstrated an innovative approach to efficiently miniaturize RIS elements while ensuring their functionality across various operational states. The integration of the Otsu image segmentation method with linear mapping and Quasi-Newton optimization algorithm has yielded optimized RIS elements with tailored structural topologies. Our experimental validation, through the design and simulation of a 16 × 16 3-bit RIS, not only confirms the high degree of accuracy between simulations and measurements but also showcases a remarkable 60\% reduction in the passive metal surface area of the RIS elements. This achievement underscores the efficacy of our proposed algorithm in offering a practical and efficient solution for the design and optimization of RIS elements, paving the way for further advancements in this emerging field.

\bibliographystyle{IEEEtran}
\balance
\bibliography{qjh_ref.bib}

\end{document}